**Articles You May Be Interested In**

Completing the dark matter solutions in degenerate Kaluza-Klein theory

*J. Math. Phys.* (April 2019)

Gibbs measures based on 1d (an)harmonic oscillators as mean-field limits

*J. Math. Phys.* (April 2018)

An upper diameter bound for compact Ricci solitons with application to the Hitchin–Thorpe inequality. II

*J. Math. Phys.* (April 2018)

# Ultrafast pulse duration measurement method of near-infrared pulses for a broad range of wavelengths using two-photon absorption in a liquid and fluorescent dye solution

Rafeeq Syed[a]  and Cornelis J. G. J. Uiterwaal

AFFILIATIONS
Department of Physics and Astronomy, University of Nebraska Lincoln, 855 N 16th Street, Lincoln, Nebraska 68588, USA

[a]Author to whom correspondence should be addressed: rafeeq@huskers.unl.edu

ABSTRACT

A novel characterization method to measure the pulse duration of ultrafast near-IR pulses is introduced, which uses simple tabletop optics, is relatively inexpensive, and is expected to work in a broad wavelength range. Our diagnostic tool quantitatively characterizes the laser pulse duration of any near-IR wavelength assuming a Gaussian pulse shape with a linear chirp. We negatively prechirp near-IR pulses with a home-built broadband pulse compressor (BPC) and send this prechirped beam through a cell filled with a low-molar solution of a fluorescent dye in a liquid. After two-photon absorption, this dye fluoresces in the visible, and we record this visible signal as a function of the propagation distance in the liquid cell. We calibrate the group velocity dispersion (GVD) of our home-built BPC device against the known GVD of the compressor of our 800 nm laser and confirm this value using geometric considerations. Now knowing the GVD of BPC and the recorded visible signal for various amounts of negative chirp, let us extract the smallest pulse duration of the near-IR pulse from this visible signal. As a useful corollary, our analysis also enables the direct measurement of the GVD for liquids and the indirect measurement of the absorption coefficient for liquids in the near-IR range, in contrast to indirect GVD measurements that rely on methods such as the double derivative of the refractive index.

## I. INTRODUCTION

Ultrashort pulses passing through optical material experience group delay dispersion (GDD), which causes them to stretch and become chirped. On the contrary, they can also be compressed if the initial pulse is negatively chirped. This is due to different wavelengths traveling at different speeds. Measuring ultrashort pulses near the Fourier transform limit (FTL) is necessary because they are more susceptible to this effect, and the pulse duration is increased dramatically. Multiple robust methods have been used to measure ultrashort pulses such as FROG,[1] SPIDER,[2] MIIPS,[3] D-scan,[4] Wizzler,[5] etc. Techniques such as Dazzler–Wizzler combo[5] and transverse SHG in disordered crystals[6] provide a real-time analysis of pulse evolution and facilitate rapid correction of pulse chirp acquired during propagation through an optical system. While most of these are available as commercial devices, are expensive, and measure complete temporal intensity and phase. Home-built autocorrelators have comparable accuracy and are cheaper. In addition, autocorrelators rely on assuming the pulse shape, but the duration is directly extracted from the autocorrelation width. All methods use nonlinear interaction, which can be sensitive to alignment, require high laser power, and use specific wavelength-specific nonlinear crystals. While alignment sensitivity holds true in most cases, d-scan[4] or amplitude swing[7] are inline techniques that employ a single-beam configuration and exhibit insensitivity to alignment. There is an increasing interest in developing reliable techniques for the precise temporal characterization of ultrashort pulses that can be applied across various spectral ranges and accommodate differ-ent pulse bandwidths.[8] The limitation based on the availability of nonlinear crystals for specific wavelengths challenges the scientific community to measure ultrafast pulses at various wavelengths with a single setup. Several strategies have been employed to mitigate crystal limitations, such as the utilization of nano d-scan.[9] In addition, FROG and the amplitude swing have been shown to cover a wide spectral range.[8,10] The goal of this paper is to introduce a pulse measurement device that is relatively inexpensive and that it is expected to work in a broad wavelength range. This new device is based on a previous experiment[11] done by our group, combined with a home-built Broadband Pulse Compressor (BPC) based on the Trebino BOA compressor.[12,13]

Below, we will describe our setup in three steps:

- First, we present the experimental setup shown in Fig. 1, which is further elaborated in Figs. 3 and 4. The setup consists of two stages. In Stage I, prepared the laser pulse by passing through BPC. In Stage II, the pulse propagates through a glass tube with a fluorescent dye solution, generating a fluorescence signal captured by a camera.
- Next, we will discuss our estimation of the GDD of the BPC using its geometrical parameters and discuss the theory of Gaussian pulse propagating in dispersive media (such as our liquids).
- Finally, we verify the experimentally measured value of the BPC's GDD against the theoretical value we estimated using its geometrical parameters.

## II. EXPERIMENTAL SETUP

We used 800 nm, femtosecond pulses from our Ti:sapphire Spitfire laser (repetition-rate 1 kHz). The beam diameter was 5 mm and has a full width half maximum (FWHM) bandwidth of 35 nm centered at 800 nm (measured with a spectrometer). This system has a grating-based pulse compressor and depending on how much initial chirp was needed, the pulse duration from the Spitfire laser system was varied from ~50 to 200 fs. The pulses passed through our BPC as shown in Figs. 3 and 4. The output from the BPC was sent through a liquid dye solution confined in a glass tube with a dimension of 304 × 14 × 14 $mm^3$. The entrance of the glass tube was a 3.25 mm thick UV silica flat window, and the other end was sealed with black Viton to reduce light scattering. We used pure water or pure methanol, in which small amounts of the dye fluorescein were dissolved. A Canon T6i camera was used to capture the fluorescence signal from the dye. Sample images captured are shown in Fig. 2.

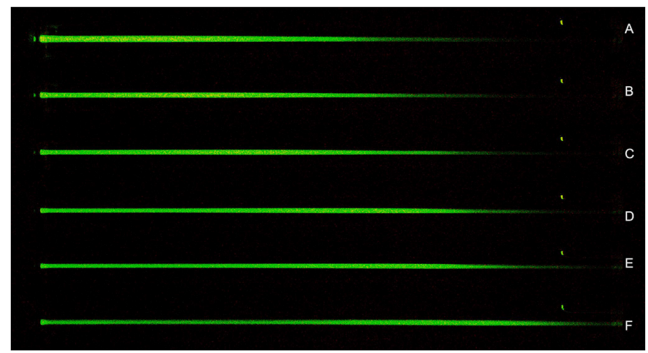

**FIG. 2.** Compilation of photographs of the fluorescence in the liquid cell containing the fluorescein dye solution. The laser beam enters the cell on the left. The amount of negative chirp in the incoming pulse is increased from A to F, which means a longer pathlength in the liquid is needed to compress the pulse. In F, the brighter signal is near the right edge, where the pulse is shortest.

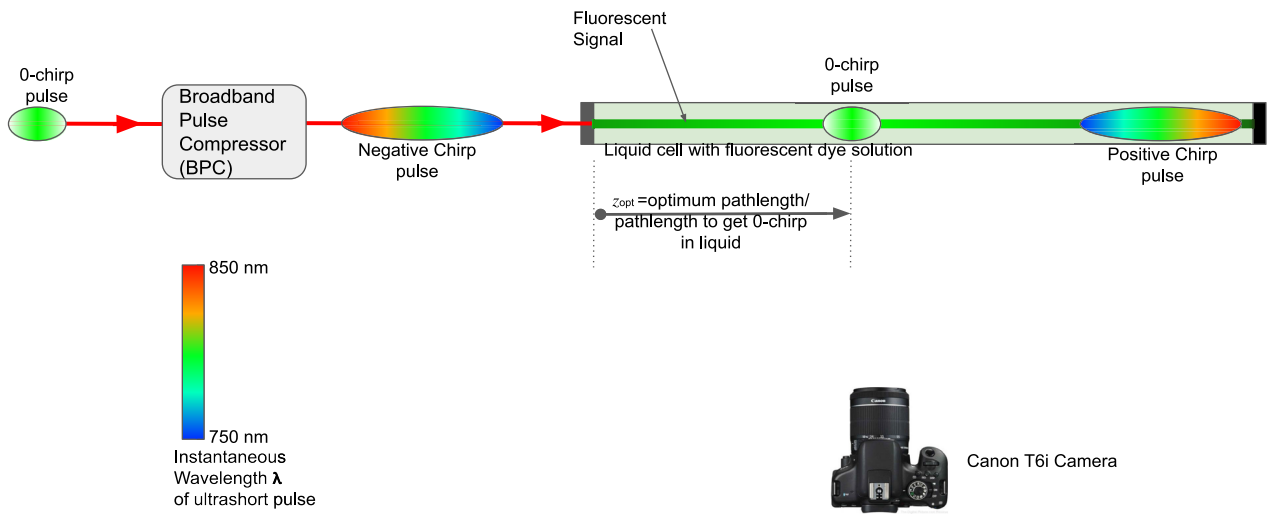

**FIG. 1.** Principle of the experiment. We send negatively chirped pulses into a liquid and gain insight into pulse duration by measuring the fluorescence signal due to two-photon absorption by the dye. This signal is proportional to intensity squared or inverse pulse duration.

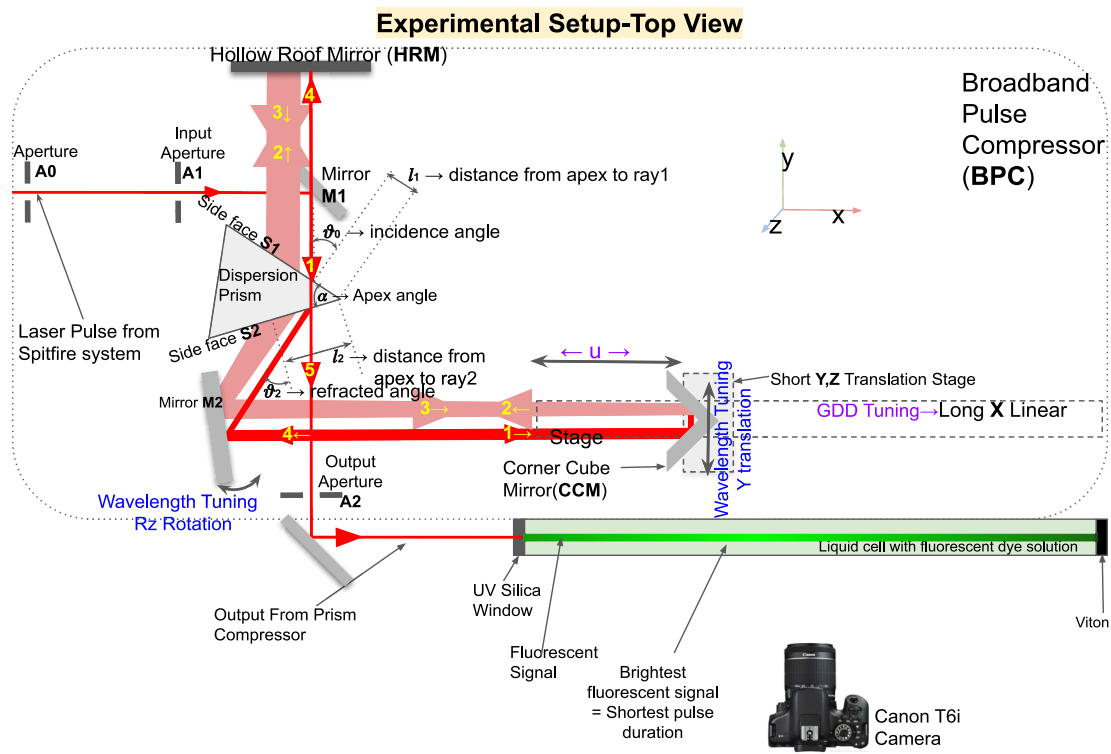

**FIG. 3.** Top view of our experimental setup. A broadband prism compressor (BPC) shown here inside a black dotted rectangle. Apertures A0 and A1 fix the input axis for the compressor. For wavelength tuning, mirror M2 is rotated about the *z* axis, as shown by the arced arrow when the rays (1, 3) are not parallel to the *X* linear stage. The Corner Cube Mirror (CCM) is translated along the *y* axis to change $l_2$ the distance from the apex to the ray (2), to avoid clipping on the prism's S2 face. Both CCM and M2 are used during wavelength tuning to bring ray (5) to output aperture A2 after the central wavelength $\lambda_0$ is changed. The position *u* of the CCM along the *X* direction is used to tune the GDD as long as ray (2) does not clip on the prism face S2. Ray (5) exits the compressor passing above the prism and enters the liquid cell with fluorescent dye solution. As the negatively chirped pulse is passed through the solution, the pulse is compressed due to the GVD of the liquid, and somewhere inside the cell we get the shortest pulse duration, which gives the brightest fluorescent signal due to higher intensity based two photon absorption and is captured by the Canon T6i camera.

## A. Broadband pulse compressor-(BPC)

We built the BPC (Fig. 3) inspired by Swamp Optics's BOA.[12,13] As shown in Figs. 3 and 4, an additional mirror M2 was placed in the compressor for wavelength tuning, such that the refracted rays (1, 3) could be directed toward the corner cube mirror (CCM). The CCM is mounted on a short *YZ* translation stage to compensate for the position shift of the rays (2, 4) returning from the CCM. These modifications increase the wavelength range of the BPC, so that a single setup can be used instead of multiple setups, which have a limited wavelength tuning range of 200 to 400 nm as the BOA compressor requires.[12] Instead of rotating the prism for wavelength tuning,[12] we tune M2 and the *YZ* stage for wavelength tuning. This tuning method also increases efficiency because the incidence angle $\vartheta_0 \sim 59.7°$ (Fig. 3) does not change and can be kept close to the Brewster angle. The prism in Fig. 3 was made of SF10, and its apex angle was $\alpha = 60.6°$. For SF10, the Brewster angle only changes by 1.33° between 400 and 2000 nm. Thus, keeping the incidence angle almost the same reduces unwanted reflections and results in high efficiency for our BPC. Even though BPC requires some minimal alignment as opposed to BOA compressors, its extended tuning range, its lower cost for that extended tuning range, and the absence of the hassle of changing BOA compressors for different central wavelengths make our BPC an attractive option.

## III. COMPRESSOR GEOMETRY
### A. Optics toolbox—Two-prism compressor

To numerically determine the GDD and third-order dispersion (TOD) of our BPC, we used a toolbox.[14,15] It determines GDD and TOD for some central wavelength $\lambda_0$ based on a prism compressor's geometry (Fig. 5), involving quantities such as incidence angle $\vartheta_0$, prism apex angle $\alpha$, distance from apex to ray 1 ($l_1$), and distance from apex to ray 2 ($l_2$). ($L/2$) is the length projection perpendicular to the incident prism face, for the length from the prism's apex to the corner of CCM.

We inserted a factor of 2 in the pathlength from prism 1 to prism 2 because the toolbox uses a two-prism compressor and our BPC is a single prism. We measured values of $\vartheta_0$, $l_1$, $l_2$, and $L$ using ImageJ software[16] by taking top-view pictures with a high-resolution camera. Multiple images were taken from various angles and positions to avoid parallax errors.

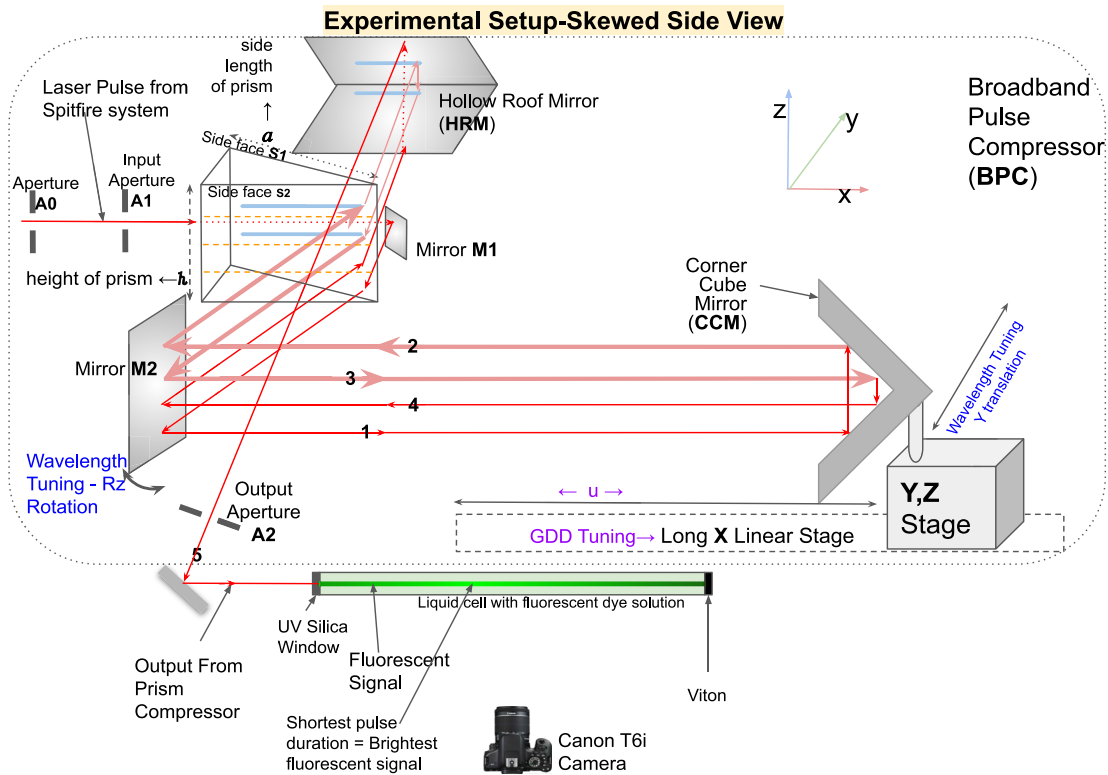

**FIG. 4.** Skewed side view of the experimental setup. This figure complements Fig. 3 by showing the vertical position of all rays in prism, on HRM, M1, M2, and CCM. Here, the spatially widened beams (2, 3) are shown as simple rays for clarity, but their horizontally stretched nature can be seen in the top view Fig. 3 and in here as the blue lines on side face S2 and HRM. The three orange dashed lines on S2 mentally divide the prism into four layered sections of $h/4$ thickness, which are stacked vertically. Here, rays pass through these prism layers in order of ray 2, 3, 4, 1 from top to bottom, and ray 5 is passing above the prism. In this setup, HRM, prism, and CCM all have a vertical size of 1 in., which constrains the place for input and exit beams. So, we had to use M1 for input ray 1 and HRM for exit rays (4, 5). Note: The corner cube mirror flips an image in the horizontal and vertical directions.

Here, GDD and TOD are given by[14]

$$\text{GDD}(\lambda) = -\frac{\lambda^3}{2\pi c^2}\frac{d^2 P}{d\lambda^2}, \quad (1)$$

$$\text{TOD}(\lambda) = -\frac{\lambda^4}{4\pi^2 c^3}\left(3\frac{d^2 P}{d\lambda^2} + \lambda\frac{d^3 P}{d\lambda^3}\right), \quad (2)$$

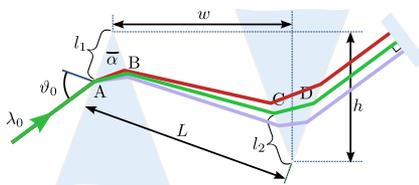

**FIG. 5.** Reproduced with permission from Light Conversion, "Two prism compressor" (2023).[18] Copyright 2023 Light Conversion, UAB.

where $P = 2[n(\lambda)P_{\text{AB}} + P_{\text{BC}} + n(\lambda)P_{\text{CD}} + P_{\text{DE}}]$. An extensive phase delay analysis for two prisms has been done in Ref. 15.

### B. Optics toolbox—Grating-pair compressor

The GDD and TOD of the Spitfire laser's grating-based compressor were numerically estimated using a grating pair compressor toolbox.[17,18] To calculate the dispersion for a specific central wavelength ($\lambda_0$), we again need geometrical quantities: the incidence angle ($\vartheta_0$), grating pitch ($d$), and the perpendicular distance between gratings ($L$). These geometrical quantities were also measured using ImageJ[16] Sec. III A. The GDD and TOD for the grating pair compressor are, respectively (Fig. 6),[17]

$$\text{GDD}(\lambda) = -\frac{\lambda^3 L}{\pi c^2 d^2}\left[1 - \left(\frac{\lambda}{d} - \sin(\vartheta_0)\right)^2\right]^{-3/2}, \quad (3)$$

$$\text{TOD}(\lambda) = -\frac{3\lambda}{2\pi c}\frac{\partial^2 \phi_g}{\partial \omega^2}\frac{1 + \frac{\lambda}{d}\sin(\vartheta_0) - \sin^2(\vartheta_0)}{1 - \left(\frac{\lambda}{d} - \sin(\vartheta_0)\right)^2}. \quad (4)$$

An extensive path delay analysis for two grating compressors has been done in Ref. 18.

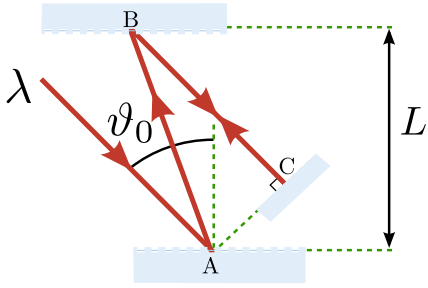

**FIG. 6.** Reproduced with permission from Light Conversion, "Grating pair compressor" (2023).[18] Copyright 2023 Light Conversion, UAB.

## IV. THEORY

From a theoretical point of view, the dispersed pulse duration $\Delta t$ is related to the transform-limited pulse duration $\Delta t_{opt}$ as[11]

$$\Delta t = \Delta t_{opt}\sqrt{1 + \left(4\ln 2\frac{\text{GDD}}{\Delta t_{opt}^2}\right)^2}, \quad (5)$$

where a Gaussian profile with a linear chirp is assumed with an FWHM of $\Delta t$. Positive GDD causes shorter wavelengths to lag behind longer wavelengths, while negative GDD causes shorter wavelengths to lead ahead of longer wavelengths. The final pulse duration depends on the magnitude of GDD. Here, we only consider linear dispersion, as second order dispersion, usually has the biggest contribution. TOD may be ignored as long as it is significantly smaller than $\Delta t_{opt}^3$.

When a pulse travels in an optical medium (om), it gathers a positive GDD depending on the pathlength $z$. The relevant quantity is the group velocity dispersion (GVD) denoted by $K_{om}$, which is related to the GDD by

$$\text{GDD}_{om} = K_{om} * z. \quad (6)$$

Thus, the GVD of an optical medium is the GDD per unit pathlength ($z$).[19] Pulse stretchers/compressors can also cause GDD, where the GDD of compressors is usually negative. This GDD of compressors is varied by changing the distance between their dispersive/diffractive optical elements. In our experiments, we adjust the GDD by moving the retro-reflector in the Spitfire's compressor and by moving the corner cube mirror in our BPC.

For the calculations that follow, we define a quantity we call the compressor parameter $G_C$ as the change in GDD per unit change in the retroreflector position $u$, which is similar to a GVD. Thus, we write

$$G_C = \frac{\Delta \text{GDD}}{\Delta u}, \quad (7)$$

where $u$ includes a factor of 2 as mentioned earlier. The compressor parameter $G_C$ is fixed by the pulse compressor's geometry.[11] As shown in Fig. 1, when a pulse with negative chirp travels through an optical medium, it will acquire zero chirp after a certain distance, which we call $z_{opt}$. This distance is given by[11]

$$z_{opt}(u) = \frac{G_C(u - u_{opt})}{K_{om}}. \quad (8)$$

Here, $u_{opt}$ is the CCM's position where the output pulse from the compressor is transform-limited, i.e., $z_{opt} = 0$. For $z_{opt} = 0$, the pulse will only get stretched by liquid and will never be transform-limited at any point in liquid. For fixed $u$, the pulse duration is changed by the liquid's GVD such that at any given position $z$ in liquid, the pulse duration is[11]

$$\Delta t(z) = \Delta t_{opt}\sqrt{1 + 16(\ln 2)^2 K_{om}^2 (z - z_{opt})^2 / \Delta t_{opt}^4}. \quad (9)$$

If pulse energy loss due to absorption by liquid is ignored, the peak intensity of the pulse at position $z$ in the liquid is[11]

$$I_{peak}(z) = \frac{I_{peak}(z_{opt})}{\sqrt{1 + 16(\ln 2)^2 K_{om}^2 (z - z_{opt})^2 / \Delta t_{opt}^4}}. \quad (10)$$

In our experimental setup, we obtain pulse durations by analyzing the fluorescence signal intensity of the dissolved dye (for instance, fluorescein). By keeping the dye concentration low, its energy absorption is insignificant, and the GVD we measure is unaffected by the dye. The dye's only purpose is to produce fluorescence in the visible after two-photon absorption (TPA) from near-IR light.

We use low-pulse intensities to avoid saturation of the dye's TPA. Based on the published TPA cross section of 54 GM,[20] we estimated that the energy absorbed by dye molecules in the entire cell is only a few percent.[21] Thus, we kept the beam power below 20 mW. Thus, recorded fluorescent light at any location $z$ in the liquid is proportional to the time-integrated square of the pulse's intensity profile. The visible signal, denoted here as $S(z)$, is (omitting proportionality factors):[11]

$$S(z) \propto \int_{-\infty}^{+\infty} I^2(t,z)dt = I_{peak}^2(z)\sqrt{\frac{\pi}{4}}\frac{\Delta t(z)}{\sqrt{2\ln 2}}$$
$$= \sqrt{\frac{\pi}{8\ln 2}}\frac{E_{tot}^2(z)}{\Delta t(z)}, \quad (11)$$

where $E_{tot}(z) = I_{peak}(z)\Delta t(z)$ is the total energy of the pulse as a function of position $z$. Absorption by the liquid decreases this energy exponentially as the pulse propagates, following the Lambert–Beer law $E_{tot}(z) = E_0 \exp(-\beta z)$, where $\beta$ is the absorption coefficient.[22] Absorption coefficients are typically recorded for continuous-wave beams, but we will assume the same absorption coefficient is valid for ultrashort pulses. The fluorescence signal in Eq. (11) is then

$$S(z) \propto \frac{\exp(-2\beta z)}{\Delta t(z)}$$
$$= \frac{\exp(-2\beta z)}{\Delta t_{opt}\sqrt{1 + 16(\ln 2)^2 K_{om}^2 (z - z_{opt})^2 / \Delta t_{opt}^4}}. \quad (12)$$

By capturing the fluorescence signal $S(z)$ from the liquid cell (Fig. 2), we can study the GVD of the liquid ($K_{om}$) and the duration of the pulse $\Delta t(z)$ as it propagates in the liquid. By measuring position $z_{opt}$ (strongest fluorescence) for various positions of the retroreflector (so for various chirps of pulse as it enters the cell), we find $K_{om}$. This is done by taking the slope of the $z_{opt}$ vs $u$ graphs. This slope equals $G_C/K_{om}$ [see Eq. (8)]. The quantity of $G_C$ only depends on the compressor's geometry. Once $K_{om}$ is found, we can estimate the transform-limited pulse duration $\Delta t_{opt}$ from the FWHM value $z_{1/2}$ of the fluorescence signal $S(z)$, after first correcting this signal for the exponential absorption $\exp(-2\beta z)$. The transform-limited pulse duration $\Delta t_{opt}$ is given by[11]

$$\Delta t_{opt}^2 = K_{om} \frac{2 \ln 2}{\sqrt{3}} z_{1/2}. \quad (13)$$

As the value of $S(z_{opt})$ does not depend on $\Delta t$ or $\Delta t_{opt}$, the value of $\beta$ can be found from an exponential decay fit to the graph of $S(z_{opt})$ vs $z_{opt}$. After obtaining the transform-limited pulse duration $\Delta t_{opt}$ in the liquid, the duration of the laser pulse itself can be estimated from Eq. (5), using the GDD caused by the BPC and the GDD caused by the liquid. The practicality of this method of estimating $\Delta t$ using GDD and $\Delta t_{opt}$ has been shown in Fig. 5 of Ref. 11.

The GDD of the BPC can be estimated using tools such as Optics toolbox[14,15] described earlier. Note that the GDD of the BPC will be negative so that the pulse entering the liquid cell has a negative chirp. The positive GDD caused by the liquid and the entrance window is estimated using Eq. (6), where the propagation length in the liquid is $z_{opt}$, the position of the maximum $S(z)$. The resultant GDD (which we will call $GDD_{all}$) is the sum of the BPC's negative GDD, the positive GDD of the liquid, and the entrance window,

$$GDD_{all} = GDD_{BPC} + K_{om} * z_{opt} + K_{window} * d, \quad (14)$$

where $K_{window}$ is the GVD of the window material (UV silica) and $d$ is its thickness. Finally, we use the measured values of $GDD_{all}$ and $\Delta t_{opt}$ in Eq. (5) to estimate the duration of the laser pulse.

### A. Bandwidth limitations

In our setup, the FWHM bandwidth $\Delta \lambda = 2.355 \sigma_\lambda$ is limited mainly by the side length S(=S1 = S2), of the prism used in BPC. To maximize the bandwidth limit, the distance $l_2$ from the apex to ray 2 as shown in Fig. 3 should be set such that ray 2 is incident exactly in the middle of side S2, i.e., $l_2 = S2/2$. The angle of refraction $\vartheta_2(\lambda_0)$ defined as the angle between ray 1 and the normal of S2 (see Fig. 3) is needed and is different based on the prism material and the central wavelength $\lambda_0$. The projection width of S2, which can be used by the dispersed beam (ray 2), is effectively $S2 \cos(\vartheta_2)$. To avoid ray 2 clipping at S2, we need the whole bandwidth $\sim 6\sigma_\lambda$ to use this effective width $S2 \cos(\vartheta_2)$. Using the thin prism approximation, we can say that the dispersion angle $\delta_{\vartheta_2}$ for the whole bandwidth $6\sigma_\lambda$ is approximately

$$\delta_{\vartheta_2} \approx 6\sigma_\lambda \times \frac{\partial n}{\partial \lambda}\bigg|_{\lambda_0} \times \alpha, \quad (15)$$

where $n$ is the refractive index and $\alpha$ is the apex angle of the prism. To avoid clipping of ray 2 at side S2, we need

$$\delta_{\vartheta_2} \times L \leq S2 \cos(\vartheta_2)$$
$$\implies \Delta \lambda \leq (2.355 \times S2 \cos(\vartheta_2)) \div \left(6L \times \frac{\partial n}{\partial \lambda}\bigg|_{\lambda_0} \times \alpha\right), \quad (16)$$

where $\Delta \lambda$ is the FWHM bandwidth and $L$ is the double pass length between the CCM and prism.

### B. Cell length limitation

The cell length $L_{cell}$ is limited by two things, where the first limitation is to have the peak signal approximately in the middle of the cell, i.e., $L_{cell}/2 = z_{opt}$ [see Eq. (8)] which is the balancing of $GDD_{BPC} \approx K_{om} L_{cell}/2$. So the longest pulse duration measured depends on the amount of prechirp possible by BPC and the magnitude of GVD to counter it in the cell. The second limitation is that $L_{cell}$ should be much smaller than the nonlinear length $L_{NL}$ of the liquid used, which is the length where the nonlinear effects due to ultrashort pulse propagation become appreciable. As described in Refs. 23 and 24, $L_{NL}$ is given by

$$L_{NL} = \frac{2\pi}{\lambda_o n_2 I} \quad (17)$$

where $I$ is intensity and $n_2$ is the nonlinear refractive index of the liquid used. Based on our calculations for the intensity we used, $L_{NL}$ was in the order of meters for both liquids we used. As $L_{cell}$ = 30.4 cm, much less than $L_{NL}$, we conclude our measurements were not affected by nonlinear effects.

## V. EXPERIMENTAL PROCEDURE AND DATA ANALYSIS
### A. Liquid solution preparation

We prepared a liquid dye solution by dissolving fluorescein dye in de-ionized water or pure methanol with a concentration of much less than 0.002M. This concentration was chosen because it minimizes absorption by the dye, which can impact experimental outcomes.[11] The dye solution was left undisturbed for 24 h to allow any suspended, undissolved dye particles to settle down. The mean power of the beam was reduced to about 20 mW[11] to avoid nonlinear absorption effects in the liquid. Methanol with 99.9% purity was obtained from Sigma-Aldrich and used as purchased.

### B. Camera selection and placement

To capture the fluorescent signal $S(z)$, we used a Canon T6i digital camera with an EF-S 18–55 mm f/4–5.6 lens. As shown in Figs. 3 and 4, the camera was placed ~1 m away from the liquid cell, focused to sharply view the signal, and tilted so that horizontal pixels were aligned with $S(z)$. A scale was placed in the focal plane to verify there are no parallax distortions. To avoid stray light, the camera and liquid cell were isolated in a dark box. Images were captured remotely with a battery adapter to avoid movement of the camera. The T6i's linear response was verified by taking multiple images with varying exposure settings. Canon's raw format CR2 was selected to ensure no image data were manipulated by the camera's processing. The data from raw CR2 files were extracted using the open-source tool dcraw.[25] We also experimented with a Nikon's D3300 camera but found that its lossless compression raw format NEF is indeed lossy, rounding intensity values near high intensities and keeping more data near low intensities. This may be sufficient for art photography but not for scientific measurements, so Canon's T6i was preferred.

### C. Data analysis

The raw image data were analyzed using MATLAB.[26] Only the green channel was used, as the fluorescence of the dye is in the green (Fig. 2). The background was subtracted, and a region of interest was selected along the fluorescent signal. $S(z)$ was taken as the mean along the vertical direction to reduce noise, and $z$ as the horizontal direction.

To fit the measured values of $S(z)$ vs $z$, we find it convenient to rewrite Eq. (12) as follows:

$$S(z) = \frac{ae^{(-bz)}}{\sqrt{1 + k(z - z_{\text{opt}})^2}} \quad (18)$$

with fit variables ($a, b, k$, and $z_{\text{opt}}$). Figure 7 shows a typical fit. The main fit variable of interest is $z_{\text{opt}}$, from which we obtain the values of $\Delta t$ and $K_{\text{om}}$.

To obtain $K_{\text{om}}$ accurately, we plot $z_{\text{opt}}$ vs $u$ for various values of $u$, the retroreflector positions. The slope of this plot is $G_C/K_{\text{om}}$ [see Eq. (8)]. The value of $G_C$ is estimated with the geometric parameters of the compressor using the procedure explained in Secs. III A and III B. Using the values of the slopes $G_C/K_{\text{om}}$ and $G_C$, we get the value of $K_{\text{om}}$. Please note that $K_{\text{om}}$ values can also be taken from available databases, such as Ref. 27, and the plot $z_{\text{opt}}$ vs $u$ is just for verification.

After obtaining the value of $K_{\text{om}}$, using fit variable $k$, we can calculate the value of $\Delta t_{\text{opt}}$ using Eq. (13) or the relation $\Delta t_{\text{opt}}^4 = 16(\ln 2)^2 K_{\text{om}}^2/k$.

Knowing the value of $\Delta t_{\text{opt}}$, we can now use Eq. (5) to obtain $\Delta t$. But we also need the value of GDD to use in Eq. (5) and we get this value by using Eq. (14).

Due to absorption, the maximum of $S(z)$ shifts from $z = z_{\text{opt}}$ to some $z = z_{\text{peak}}$. To estimate this shift, we solve for $S'(z) = 0$, finding

$$z_{\text{peak}} = z_{\text{opt}} - \frac{1}{2b} + \frac{1}{2b}\sqrt{1 - \frac{4b^2}{k}}. \quad (19)$$

From our fits, we find that, values of $b$ and $k$ for both liquids (water and methanol) are roughly $b \approx 10^{-4}$ px$^{-1}$ and $k \approx 10^{-6}$ px$^{-2}$. These values allow us to use a binomial approximation

$$z_{\text{peak}} = z_{\text{opt}} - \frac{b}{k}. \quad (20)$$

Since $b$ and $k$ are constants, the peak position has a constant shift of ~100 pixels (≈5.3 mm). This should not affect the value of $K_{\text{om}}$, the GVD of liquids obtained from Figs. 8 and 9, as the slope $G/K_{\text{om}}$ does not depend on this constant shift. Ideally, a liquid with no absorption in a given wavelength range should be preferred so that $b/k$ is small enough to be ignored. The $z_{\text{opt}}$ obtained from the fit in Fig. 7 is already corrected for the shift $b/k$. Now, using the values of $\Delta t_{\text{opt}}, K_{\text{om}}$, and $z_{\text{opt}}$, we can estimate pulse duration $\Delta t$ by utilizing Eqs. (5) and (14).

### D. Next research direction

Our current approach provides a coarse pulse duration estimation, accounting for dispersion up to the second order only. We suspect the small bump at ~4500 pixels in Fig. 7 is due to higher-order phase distortions, which will be investigated in future research.

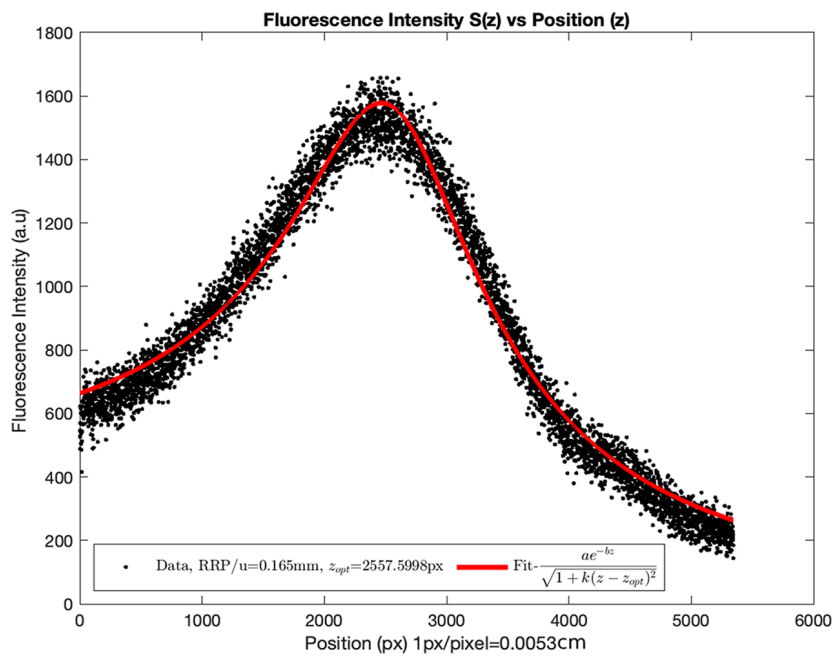

**FIG. 7.** Black dots: typical fluorescence signal $S(z)$ for a fixed value of $u$. Red curve: fit of Eq. (18) done in MATLAB to extract the position $z_{\text{opt}}$ of the transform-limited pulse in the liquid cell. When the retro-reflector position $u$ is changed, the position $z_{\text{opt}}$ changes accordingly. See Eq. (8).

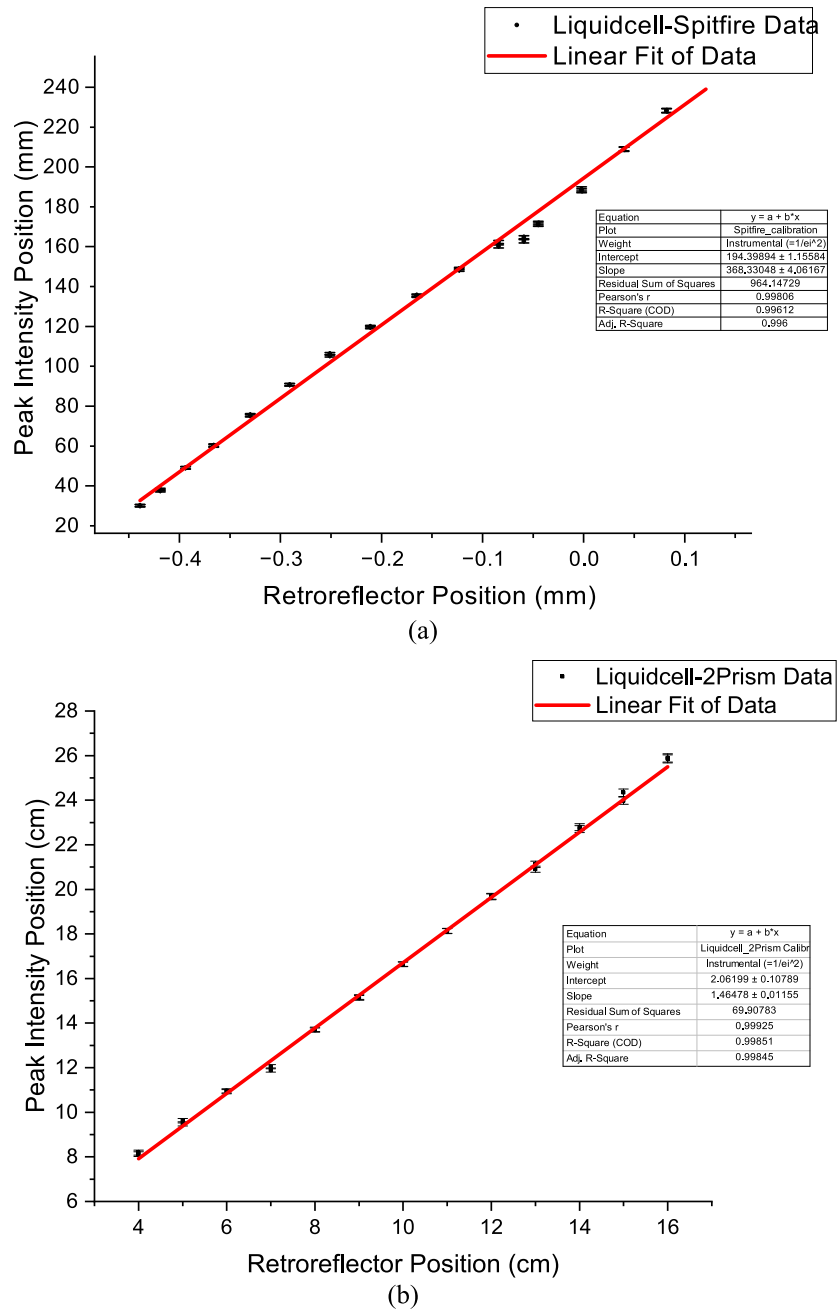

**FIG. 8.** Measurements with fluorescein-water solution. (a) Location $z_{opt}$ of peak fluorescent intensity as a function of the position $u$ of the grating compressor of the Spitfire laser system. In this measurement, the liquid cell was filled with water and fluorescein dye. (b) Location $z_{opt}$ of peak fluorescent intensity as a function of the position $u$ of the corner cube mirror of the BPC system. In this measurement, the liquid cell was filled with water and fluorescein dye.

Another direction is to spectrally resolve the fluorescent signal to obtain a two dimensional trace similar to SHG FROG or single-shot d-scan,[28] where we can possibly get full pulse information including higher-order phase distortions. The SHG signal can be analyzed to get full pulse information without ambiguity, as there are no intermediate states during the emission part of SHG. This is not the case in TPA, as the emission spectrum depends on the radiative channels and may lose the original spectrum information.

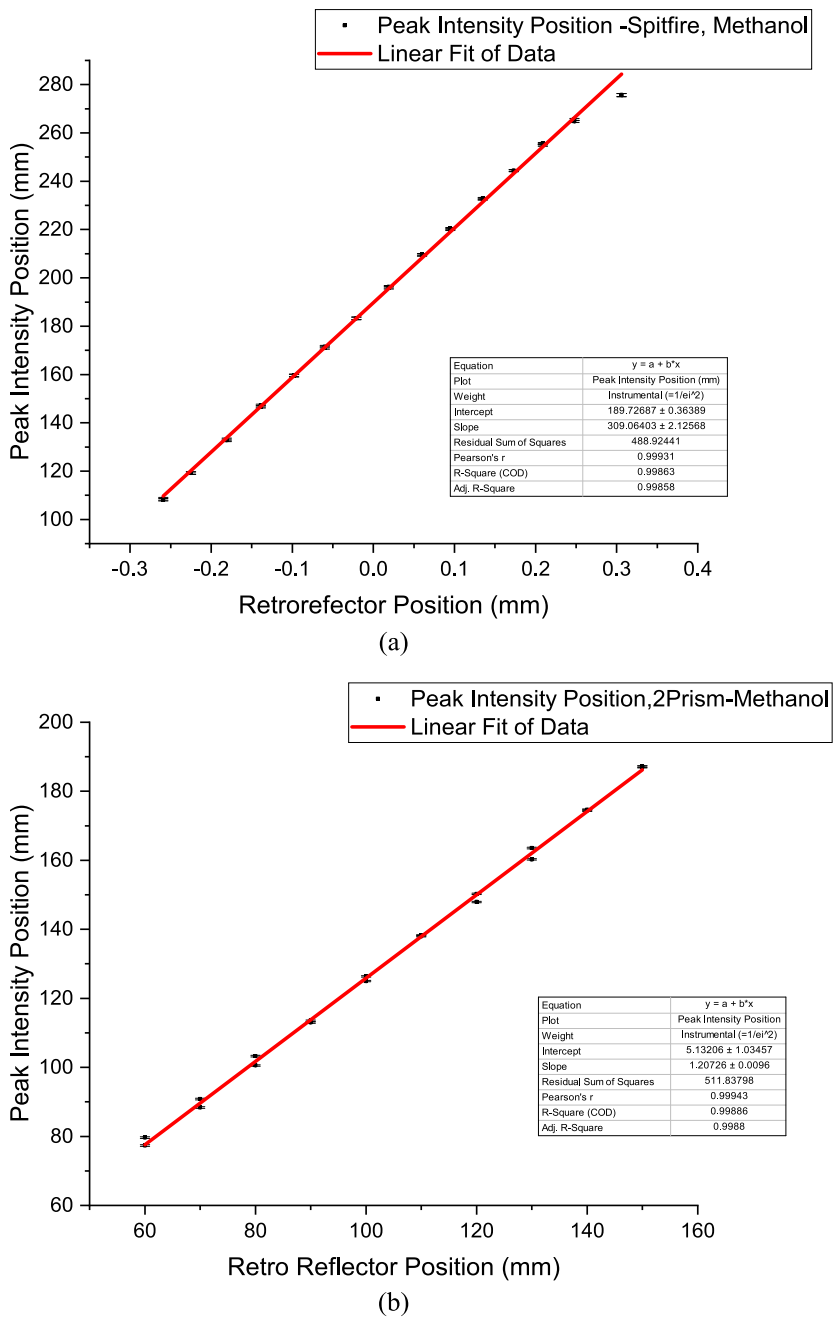

**FIG. 9.** Measurements with fluorescein-methanol solution. (a) Same as Fig. 8(a) but for methanol and fluorescein dye, and (b) same as Fig. 8(b) but for methanol and fluorescein dye.

The complexity to extract full pulse information from a spectrally resolved fluorescent TPA signal in our method is two fold. First TPA spectrum of fluorescent dye aligns with the single photon excitation spectrum only for a few dyes such as Cumarin 540A (see Fig. 20 in Ref. 21). Assuming this condition holds for the dye, the next complexity is that the fluorescent emission has a Stoke's shift and is correlated with the excitation spectrum via a mirror image rule.[29,30] This will be investigated in future research.

## VI. RESULTS

Using geometric parameters (see Secs. III A and III B), the values of the compressor constant $G_C$ at 800 nm were measured to be $(10\,800 \pm 200)$ and $(43 \pm 1)$ fs$^2$/mm for the Spitfire's Grating Compressor (SGC) and prism-based BPC, respectively.

The optical medium for the graphs in Fig. 8 is water, and Fig. 9 is methanol. The graphs of $z_{opt}$ vs $u$ are shown in Figs. 8(a) and 9(a). It follows from Eq. (8) that the slope of the linear fits in these graphs is $G_{SGC}/K_{om}$, where $K_{om}$ is the GVD of respective optical medium used. Similarly, in Figs. 8(b) and 9(b), we have plotted $z_{opt}$ vs the corner cube position $u$ of the BPC, where the slope of the linear fits in these graphs is $G_{BPC}/K_{om}$. The ratio $R$ of the slopes in Figs. 8(a) and 8(b) or the ratio of the slopes in Figs. 9(a) and 9(b) is given by

$$R = \left(\frac{G_{BPC}}{K_{om}}\right) \div \left(\frac{G_{SGC}}{K_{om}}\right) = \frac{G_{BPC}}{G_{SGC}}. \quad (21)$$

As a key point, Eq. (21) shows that this ratio $R$ is independent of $K_{om}$, the GVD of the optical medium used. Indeed, the values of $R$ we measured with water and with methanol agree (row 3 in Table I). We use this measured value of $R$ to calibrate the $G_{BPC}$ of our home-built device against the Spitfire's compressor by simply multiplying $R$ with the known value of $G_{SGC}$ (row 4 in Table I). The resulting values of $G_{BPC}$ are given in row 6 of Table I. For further confirmation, we also estimated the value of $G_{BPC}$ based only on its geometric parameters (row 5 in Table I). The values of $G_{BPC}$ measured with two different liquids and estimated from our device's geometry are all in agreement. We have demonstrated estimation of $\Delta t$ using $K_{om}$ and $\Delta t_{opt}$ in Fig. 5 of Ref. 11, by comparing various pulse durations from SGC with FROG measurements using the same procedure.

The data we measured also allow us to calculate the GVD of the two liquids (at 800 nm): $K_{om} = R * G_{SGC}$ [see Eq. (21)]. For water, we find $K^{800}_{water} = (29.4 \pm 0.9)$ fs$^2$/mm, which is in agreement with the value $K^{800}_{water} = 30.36$ fs$^2$/mm determined from the empirical formula for the refractive index of water reported in Ref. 31. For methanol, we find $K^{800}_{methanol} = (35 \pm 1)$ fs$^2$/mm. The reported values of this quantity fall in a wide range. For instance, the empirical formula for the refractive index of methanol reported in Ref. 32 implies $K^{800}_{methanol} = 168.76$ fs$^2$/mm, while a much smaller value of $K^{800}_{methanol} = 30.0$ fs$^2$/mm is reported in Ref. 33.

Note that GVD values obtained through the utilization of Eq. (8) provide direct measurements of GVD in the near-IR. In contrast, many preceding studies[31–33] have employed indirect methods, including extrapolation of the refractive index through empirical formulas or the use of the double derivative of the refractive index. However, these approaches are more vulnerable to error propagation due to inaccuracies in refractive index measurements across different wavelengths, as the changes in refractive index are relatively minor.

## VII. CONCLUSION

We presented a novel, inexpensive instrument to determine the coarse duration of ultrafast pulses in the near-IR. It combines a home-built BPC with a liquid cell. Our instrument makes use of two-photon absorption of near-IR radiation from a dissolved dye. We determined the GVD of our instrument using its geometric parameters. This is, in principle, a single-shot device, but averaging multiple shots may be useful to increase the signal strength. We confirmed this GVD value experimentally by calibrating our instrument against the known GVD of our Spitfire laser's compressor. The feasibility of estimating $\Delta t$ using total GDD and $\Delta t_{opt}$ has been demonstrated by us in Ref. 11 by comparing $\Delta t$ to FROG measurements. In the current paper, we show the correction that needs to be done to the $z_{opt}$ due to the shift of the peak position caused by the linear absorption of the liquid. In principle, our instrument works for any near-IR wavelength (provided a suitable liquid and dye combination exists).

**TABLE I.** Summary of measured values, including the compressor constants of the spitfire grating compressor $G_{SGC}$, and the broadband pulse compressor $G_{BPC}$ determined from their measured geometries. $G_{BPC}$ is also measured experimentally independent of the BPC's geometry, using $G_{SGC}$ and using the slope ratios in row 3. The highlighted values of $G_{BPC}$ indicate that the experimental (row 6) and geometry-based (row 5) values agree.

| Row no. | Description | Water + fluorescein | Methanol + fluorescein |
|---|---|---|---|
| 1 | Peak intensity position vs Spitfire retro-reflector position slope (no units) | $368.330\,5 \pm 4.061\,7$ Fig. 8(a) | $309.064\,03 \pm 2.125\,6$ Fig. 9(a) |
| 2 | Peak intensity position vs CCM position in the BPC position slope (no units) | $1.464\,78 \pm 0.011\,55$ Fig. 8(b) | $1.207\,26 \pm 0.009\,6$ Fig. 9(b) |
| 3 | Ratio $R$ of slopes, row 2/row 1 (no units) | $0.003\,98 \pm 0.000\,05$ | $0.003\,91 \pm 0.000\,04$ |
| 4 | $G_C$ of a Spitfire's grating compressor based on its geometry | $(10\,800 \pm 200)$ fs$^2$/mm | ⋯ |
| 5 | $G_C$ of a BPC prism compressor based on its geometry | $(43 \pm 1)$ fs$^2$/mm | ⋯ |
| 6 | $G_C$ of BPC compressor $G_{BPC}$ = row 3 ∗ row 4 | $(43 \pm 1)$ fs$^2$/mm | $(42.2 \pm 0.9)$ fs$^2$/mm |
| 7 | GVD of liquids at 800 nm $K_{om}$ = row 4/row 1 | $K^{800}_{water} = (29.3 \pm 0.6)$ fs$^2$/mm | $K^{800}_{methanol} = (34.9 \pm 0.7)$ fs$^2$/mm |
| 8 | GVD of liquids at 800 nm $K_{om}$ = row 5/row 2 | $K^{800}_{water} = (29.4 \pm 0.7)$ fs$^2$/mm | $K^{800}_{methanol} = 35.6 \pm 0.9$ fs$^2$/mm |

## ACKNOWLEDGMENTS

The authors wish to thank Dr. Herman Batelaan for kindly providing us with the prism and other optics needed for the BPC, the Broadband Pulse Compressor. We also thank Joshua Beck for assistance with the operation and maintenance of our laser system.

## AUTHOR DECLARATIONS
### Conflict of Interest

The authors have filed a US Provisional Patent Application No. 63/510,841 on the technology described in this article with NUtech Ventures, an affiliate of the University of Nebraska. The authors declare no other conflicts of interest.

### Author Contributions

**Rafeeq Syed**: Conceptualization (equal); Data curation (equal); Formal analysis (equal); Investigation (equal); Methodology (equal); Resources (equal); Software (equal); Validation (equal); Visualization (equal); Writing – original draft (lead); Writing – review & editing (equal). **Cornelis J. G. J. Uiterwaal**: Conceptualization (equal); Data curation (equal); Formal analysis (equal); Funding acquisition (equal); Investigation (equal); Methodology (equal); Project administration (equal); Resources (equal); Software (equal); Supervision (lead); Validation (equal); Visualization (equal); Writing – original draft (supporting); Writing – review & editing (equal).

## DATA AVAILABILITY

The data that support the finding of this study is with in this article. The raw datasets are available from the corresponding author upon reasonable request.